\documentclass{article}[12pt]
\usepackage{epsf,amssymb,amsmath,latexsym}
\usepackage{tabularx}
\usepackage[dvips]{color,graphicx}

\begin{document}

\thispagestyle{empty}

\rightline{DAMTP-2007-45} \rightline{\tt{arXiv:0705.3363 [gr-qc]}}

\bigskip

\begin{center}

{ {\sc ESSAY ON GRAVITATION}}\footnote{Honorable Mention in the
Gravity Research Foundation Essay Competition, 2007}

\bigskip

{\large \bf The Holographic Interpretation of Hawking Radiation}

\bigskip
\bigskip
\bigskip

{ \sc Alessandro Fabbri\footnote{\tt afabbri@ific.uv.es}}

{ \it Departamento de F\'{\i}sica Te\'orica and IFIC, Universidad de
Valencia-CSIC, C. Dr. Moliner 50, Burjassot-46100, Valencia, Spain.}

\vspace{3ex}

{\sc Giovanni Paolo Procopio\footnote{\tt
g.p.procopio@damtp.cam.ac.uk}}

{ \it D.A.M.T.P., Centre for Mathematical Sciences, University of
Cambridge, Wilberforce Road, Cambridge CB3 0WA, U.K.}

\vspace*{1.5cm}

\large{\bf Abstract}
\end{center}
\noindent Holography gives us a tool to view the Hawking effect from
a new, classical perspective. In the context of Randall-Sundrum
braneworld models, we show that the basic features of
four-dimensional evaporating solutions are nicely translated into
classical five-dimensional language. This includes the dual bulk
description of particles tunneling through the horizon.

\bigskip
\bigskip

\newpage

\setcounter{page}{1}

\noindent The problem of understanding why and how black holes emit
particles is not an easy one. It is intriguing that the motivation
that led to the discovery of Hawking radiation came from classical
considerations, namely the analogy between the laws of black hole
mechanics and the laws of thermodynamics. However, already on
dimensional grounds one realizes that in order to construct a
quantity with the dimensions of an entropy starting from the area of
the event horizon $A_\mathrm{H}$, quantum mechanics needs to enter
into the game. Indeed, the only possibility is $S\sim
k_\mathrm{B}A_\mathrm{H}/A_\mathrm{Pl}$, where $A_\mathrm{Pl}$ is
Planck area and $k_\mathrm{B}$ Boltzmann's constant. This, in turn,
implies that the black hole temperature $T$ is $O(\hbar)$ and,
correctly, vanishes in the classical limit.

The actual derivation of particles emission by black holes
\cite{Hawking:1974rv} requires the quantization of matter fields in
the dynamical background describing the formation of a black hole
via gravitational collapse. This is the crucial feature allowing the
initial vacuum state of our radiation fields to be detected, by
asymptotic observers in the future asymptotic region at late times,
as a mixed thermal state at the Hawking temperature (for a
Schwarzschild black hole of mass $M$ and with $c=G=1$)
\begin{equation}\label{tempe}
    T_\mathrm{H}=\frac{\hbar}{8\pi M k_\mathrm{B}} \ .
\end{equation}

A visual mechanism useful to understand the reason for the quantum
instability of the Schwarzschild spacetime was first proposed in
\cite{HawSA}. Creation of particle-antiparticle virtual pairs in the
near horizon region may be such that the negative-energy
antiparticle is absorbed by the black hole, leaving its
positive-energy partner free to escape to infinity. Estimates for
this process indicate that the emitted particle has the correct
energy-temperature dependence we expect for thermal radiation, i.e.
$ E \sim k_\mathrm{B} T_\mathrm{H} $.

An alternative way to see this process is that of a particle
tunneling through the horizon. A recent derivation of the Hawking
effect using this idea was performed in \cite{par}, where it was
noticed that in order for the particle to tunnel through a `real
barrier' energy conservation must be enforced. Indeed, once a
particle of energy $E$ escapes from the black hole and reaches
infinity the black hole must shrink in size from $2M$ to $2(M-E)$.
Computation of the emission rate at leading order in the particle
energy $E$ gives exactly the Boltzmann factor for thermal radiation
at the Hawking temperature $T_\mathrm{H}$.

The idea of energy conservation takes us back to the real time
dependent gravitational collapse problem. Hawking's derivation, in
fixed background approximation, implies that the black hole radiates
with luminosity
\begin{equation}\label{Lumi}
    L \sim \frac{(k_\mathrm{B} T_\mathrm{H})^2}{\hbar} \sim \frac{\hbar}{M^2}
\end{equation}
for an infinite amount of time. This is at odds with the fact that
the black hole has a finite mass $M$. The classical background
spacetime itself has to be modified by the quantum corrections and
in particular the black hole mass will reduce at a rate given by
$L$. In the near-horizon region, a good approximation to the
evaporating solution is given by the advanced Vaidya metric
\cite{bardeen:1981}
\begin{equation}\label{Vai}
    ds^2=-\left(1-\frac{2M(v)}{r}\right)dv^2+2drdv+r^2d\Omega_2^2 \ ,
\end{equation}
where the apparent horizon $r_\mathrm{AH}=2M(v)$ recedes according
to
\begin{equation}\label{massre}
    \frac{dr_\mathrm{AH}}{dv}=-2L \ .
\end{equation}
Because of the evaporation apparent and event horizons, coincident
for the static Schwarzschild solution, separate. In the large mass
limit one can estimate the location of the event horizon by
integrating \eqref{massre} over the typical timescale of the process
$\Delta v  \sim M$ to get
\begin{equation}\label{eveh}
    \frac{r_\mathrm{EH}-2M}{2M} \sim -L \ .
\end{equation}
Therefore, a new region ($r_\mathrm{EH} < r < r_\mathrm{AH} $),
called in \cite{York:1983fx} the ``quantum ergosphere'' and absent
in the classical solution, forms.  In this region photons are
locally trapped but being outside the event horizon they can cross
the apparent horizon at a later time and propagate to infinity. One
can then view the (classically forbidden) tunneling trajectory of
the particle across the horizon as the (physically allowed)
trajectory across the apparent horizon in a quantum corrected
evaporating spacetime.

We propose here a new interpretation of the Hawking effect which is
entirely classical and which, nevertheless, is able to capture all
the features here described. We shall use holography, in particular
the predicted duality, in Randall-Sundrum braneworld models, between
a classical theory in $AdS_5$ (the bulk) and a theory living on our
4D universe (the brane) where classical gravity is coupled to
quantum matter fields. At linear level, the zero mode of the 5D
gravitons, bound to the brane, reproduces 4D Newtonian gravity at
large distances, whereas massive modes, propagating in the bulk,
induce the quantum effects on the brane and thus are dual to the
quantum matter fields \cite{Duff:2000mt}.

Application of these ideas beyond the linear level has led to the
conjecture \cite{conj} that for large masses ``4D black holes
localized on the brane found by solving 5D Einstein equations in
$AdS_5$ are quantum corrected black holes and not classical ones''.
Basing our arguments on this conjecture we can then describe the
Hawking effect from a new, classical prospective.

The braneworld analog of the Schwarzschild black hole is the
Randall-Sundrum Schwarzschild black string
\begin{equation}\label{efv}
ds^2=e^{-2k|z|}\left[-\left(1-\frac{2M}{r}\right)dv^2 + 2dvdr +
r^2d\Omega^2\right]+dz^2 ,
\end{equation}
where $k$ is the inverse of the $AdS_5 $ length. This metric gives
the classical Schwarzschild spacetime on the brane ($z=0$), but it
is problematic for several reasons. This solution is singular at the
$AdS $ horizon $z = \infty $ and, moreover, it is unstable to linear
perturbations (the Gregory-Laflamme instability
\cite{Gregory:1993vy}), implying a time dependent evolution away
from \eqref{efv}. If the holographic conjecture is correct, the
actual time-dependent solution will reproduce a quantum corrected
evaporating black hole on the brane. Unfortunately such solution is
not known. However, as suggested from the linear analysis, the
quantum effects on the brane are classically given by those
gravitational waves (bulk Kaluza-Klein massive modes) which
propagate along null geodesics in the bulk.

In braneworld there are two different definitions of horizons: the
brane apparent horizon $r^\mathrm{AH}_\mathrm{brane}$, defined with
respect to photons moving along null geodesics on the brane, and the
bulk apparent horizon $r^\mathrm{AH}_\mathrm{bulk}$, defined with
respect to gravitons that propagate in the full 5D spacetime.
In the
case of \eqref{efv} we have two different families of null geodesics
\cite{Chamblin:1999by}, those restricted to $z=\hbox{constant}$ (in
particular on the brane $z=0$) and those propagating nontrivially in
the bulk satisfying
\begin{equation}\label{noge}
    \dot z= -\frac{1}{k \lambda}
\end{equation}
where $\lambda $ is the affine parameter. Intuitively, one would
think that the first family of geodesics at $z=0$ defines the
brane apparent horizon, while the second one \eqref{noge} gives
the bulk apparent horizon. However, this is not so.

The first family ($z=\hbox{constant}$), equivalent to the usual
Schwarzschild null geodesics, defines the horizon at $r=2M$ where
$dr/dv=~0$. For the second family \eqref{noge} we find that
$r=2M$ is not a solution and, also, that in the near
horizon region \cite{Fabbri:2007kr}
\begin{equation}\label{drdv}
\frac{dr_\mathrm{bulk}}{dv}\sim -\frac{1}{(4Mk)^2}e^{2kz} \ , \qquad
r_{\mathrm{bulk}}(z) \sim 2M\left[1-\frac{e^{2kz}}{(4Mk)^2}\right] \
.
\end{equation}
If we now project \eqref{drdv} to the brane ($z=0$) and use the
holographic relation $\frac{1}{k^2} \sim \hbar N^2$, where $N^2$
corresponds to the (huge) number of degrees of freedom in the dual
theory, we get
\begin{equation}\label{HR}
\left.\frac{dr_\mathrm{bulk}}{dv}\right|_{0} \sim - \frac{\hbar
N^2}{M^2} \equiv -2 L \ , \qquad \frac{r_{\mathrm{bulk}}(0)-2M}{2M}
\sim -L \ .
\end{equation}
The similarity with \eqref{massre} and \eqref{eveh} is surprising.
What is the physical meaning of the surface $r_{\mathrm{bulk}}(z)$?
We cannot identify it
%$r_{\mathrm{bulk}}(z)$
with the bulk apparent
horizon, as for the black string
$r^\mathrm{AH}_\mathrm{brane}=r^\mathrm{AH}_\mathrm{bulk}=2M$.

Note, however, that the discovery of the Hawking effect was
performed in fixed background approximation and that it implied that
the Schwarzschild solution is quantum mechanically unstable and must
be modified by the quantum corrections. In a similar fashion, if the
holographic interpretation \eqref{HR} is correct it is reasonable to
expect that bulk massive modes, traveling along the nontrivial bulk
geodesics \eqref{noge} and responsible for the quantum effects on
the brane, will modify the black string to a time dependent solution
with the horizon satisfying the dual behavior \eqref{HR} on the
brane.
%In addition to this, it was shown in [] that in time dependent
%braneworld black hole solutions bulk and brane apparent horizons
%are generically distinct and in particular one always has
%$r_\mathrm{bulk }^\mathrm{AH }<
%r_\mathrm{brane}^\mathrm{AH}$.

Following this line of reasoning, the surface $r_{\mathrm{bulk}}(z)$
will become the bulk apparent horizon
$r_{\mathrm{bulk}}^\mathrm{AH}(z)$ of such time dependent solution.
This is somehow confirmed by the fact that, as shown in
\cite{Shiromizu:2000pg}, in time dependent braneworld black hole
solutions brane and bulk apparent horizons are generically distinct
and in particular $r_\mathrm{bulk }^\mathrm{AH }<
r_\mathrm{brane}^\mathrm{AH}$. Moreover, according to \eqref{HR} the
projection on the brane of $r_{\mathrm{bulk}}(z)$ will play the role
of the event horizon in the dual evaporating theory. Gravitons can
be emitted from the region $r>r_{\mathrm{bulk}}^\mathrm{AH}$ on the
brane to the bulk, and in particular from the (large) quantum
ergosphere
 $r_\mathrm{bulk }^\mathrm{AH }<r<
r_\mathrm{brane}^\mathrm{AH}$ (large because its size is
proportional to $N^2$). This is indeed the region from where we
expect the bulk massive modes dual to the Hawking quanta to be
emitted.
%This is indeed the region we expect bulk massive modes to be emitted
%which are dual to the Hawking quanta.
% will then form on the brane, the emission
%of gravitational waves from this region into the bulk being the
%likely classical process dual to Hawking radiation.

\begin{figure}[htb]
\centering \includegraphics[angle=0,width=5in] {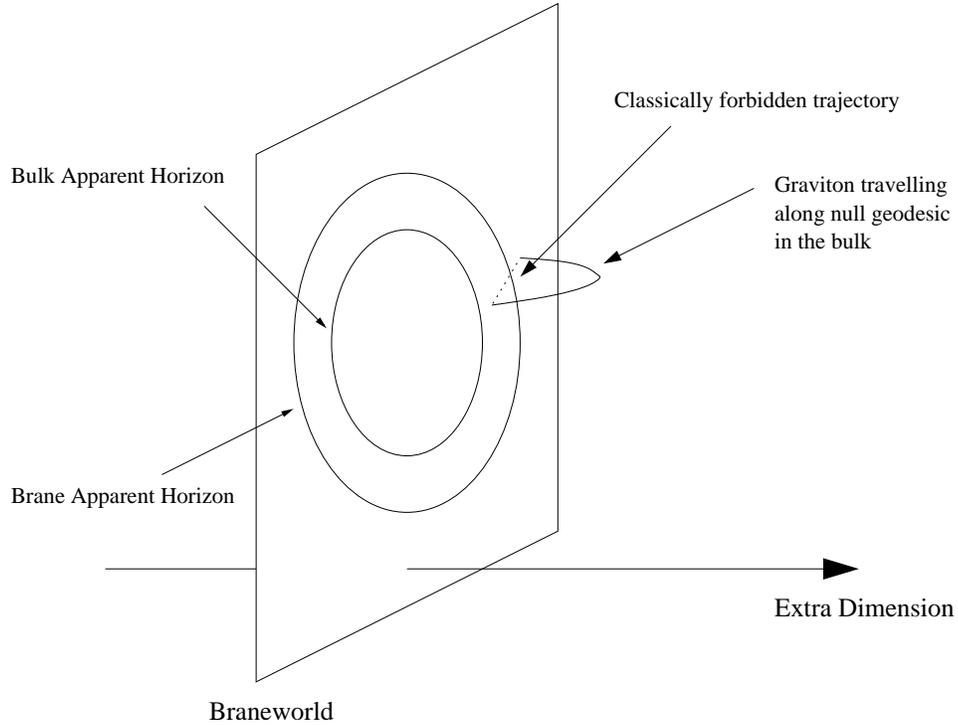}
\caption{\label{uno}  Classical bulk dual of particles tunneling
through the horizon.
%In the dual theory,  modes of
%the quantum fields `disappear' from inside the apparent horizon and
%`reappear' just outside.
}
\end{figure}

%Finally, our scenario offers the possibility described
On the base of this interesting analogy between classical bulk and
quantum brane effects we propose a mechanism that represents the
dual bulk description of particles tunneling through the horizon.
This is represented in fig. \ref{uno},
in which bulk massive modes leaving the brane
%from the
%quantum ergosphere
to the bulk bounce back to the brane just outside the brane apparent
horizon.
%Gravitons traveling along null geodesics in the bulk are dual to
%quantum modes on the brane that `disappear' from inside the apparent
%horizon and `reappear' just outside.
Whether this process really takes place or not in the full
time-dependent 5D solution could then tell us if tunneling is the
actual process behind the Hawking effect.
%\newpage

\end{document}